\documentclass[journal,12pt,onecolumn]{IEEEtran}

\usepackage[left=2cm,right=2cm,top=2cm,bottom=2cm,bindingoffset=0cm]{geometry}

\usepackage{amssymb}
\usepackage{graphicx}

\newcommand{\R}{\mathbb{R}}
\newcommand{\Z}{\mathbb{Z}}

\title{Semi-Fragile Image Authentication based on CFD and 3-Bit Quantization}

\author{
\IEEEauthorblockN{Aleksey Zhuvikin$^{\#1}$, Valery Korzhik$^{\#2}$, Guillermo Morales-Luna$^{*3}$} \\ \ \\
\IEEEauthorblockA{$^\#$Department of Secured Communication Systems, \\ The Bonch-Bruevich Saint Petersburg State University of Telecommunications, \\ Saint-Petersburg,  Russia\\
$^1$zhuvikin@ya.ru, $^2$val-korzhik@yandex.ru}
\\ \ \\
\IEEEauthorblockA{$^*$Computer Science\\
CINVESTAV-IPN\\
Mexico City, Mexico\\
$^3$gmorales@cs.cinvestav.mx}
}

\begin{document}
\maketitle              % typeset the title of the contribution

    \begin{abstract}
There is a great adventure of watermarking usage in the context of conventional authentication since it does not require additional storage space for supplementary metadata. However JPEG compression, being a conventional method to compress images, leads to exact authentication breaking. We discuss a semi-fragile watermarking system for digital images tolerant to JPEG/JPEG2000 compression. Recently we have published a selective authentication method based on Zernike moments. But unfortunately it has large computational complexity and not sufficiently good detection of small image modifications. In the current paper it is proposed (in contrast to Zernike moments approach) the usage of image finite differences and 3-bit quantization as the main technique. In order to embed a watermark (WM) into the image, some areas of the Haar wavelet transform coefficients are used. Simulation results show a good resistance of this method to JPEG compression with $\mbox{\rm CR}\leq 30\%$ (Compression Ratio), high probability of small image modification recognition, image quality assessments $\mbox{\rm PSNR}\geq 40$ (Peak signal-to-noise ratio) dB and $\mbox{\rm SSIM}\geq 0.98$ (Structural Similarity Index Measure) after embedding and lower computation complexity of WM embedding and extraction. All these properties qualify this approach as effective.
\end{abstract}


\section{Introduction}

\IEEEPARstart{I}{n} the Future Generation Services, it is very important the copyright protection in digital media such as motionless images, videos, audio files and so forth. Several methods are well known for authentication within cryptography, e. g. digital signature (DS)~\cite{i}. But if DS is appended to the work it can be corrupted or even lost in ordinary usage. If strong authentication is required, it is necessary to add an extra size of about 1,000 bits~\cite{i}. Moreover, it is quite common to compress digital images by means of the JPEG/JPEG2000 algorithms and then verification of the appended DS can be broken. The way out of this situation is to embed the DS into digital images (DI) using a watermarking technique. The first problem in the design of WM-based authentication is to build the so-called {\em reversible WM} where DI can be reversed exactly after correct WM extraction. Such technique has been proposed in many papers~\cite{ii,iii,iv,v}. But in our scenario with possible JPEG image compression it is required, in addition, that authentication be tolerant to such transform and simultaneously be able to detect any modification of the image content. In contradiction with the exact authentication such one is called \emph{semi-fragile watermarking}~\cite{vi}.

An attempt to solve this problem by means of such functionals as Zernike moments has been undertaken in~\cite{vii}, and other methods for semi-fragile or selective authentication were considered in~\cite{viii,ix,x,xi,xii}. Recently, an useful idea of~\cite{vii} concerning the usage of Zernike moments, adding the theoretical proof of the required properties and confirming the main results by simulation was presented in~\cite{xiii}. Unfortunately, last method requires a rather complex calculation of the Zernike moments that increase significantly the time of WM embedding. In the current paper, there are proposed other DI functionals, namely \emph{central finite differences} (CFD)~\cite{xiv}, which are very sensitive to  modifications of image content but simultaneously they are tolerant to JPEG compression with joint use of the so called {\em 3-bit quantization}. The calculation of CFD is much simpler than the calculation aiming to find Zernike moments, hence a reduction of the WM embedding time is obtained. Moreover, as experiments show that the new method allows to detect even small content modification better than the method considered in~\cite{xiii}.

Since any color image can be presented as few gray-scaled channels, the algorithm can be applied either to each of the channels or to both of them independently. Let assume that, for images in the RGB-model, a WM can be embedded only within the blue-channel component due to peculiar properties of \emph{Human Visual System} (HVS). Due to these facts, we will examine only grayscale images.

The reminder of the article is organized as follows. Section~\ref{sc.II} of the paper presents the main properties of CFD and the explanation of a hashing with help of 3-bit quantization method. In Section~\ref{sc.III}, the embedding and extraction algorithms are considered. The simulation results are presented in Section~\ref{sc.IV}. We conclude the paper in Section~\ref{sc.V}.

\section{The main properties of CFD and description of hashing %\\
 using 3-bit quantization} \label{sc.II}

{\em Central finite differences} (CFD) of the first order is function $\{0,1,\ldots,n_x\}\times\{0,1,\ldots,n_y\}\to\Z^+$ that is defined~\cite{xiv} as:
\begin{eqnarray}
\delta_x(x,y) &=&  \frac{1}{2}\left(I(x+1,y)-I(x-1,y)\right),  \label{eq.01} \\
\delta_y(x,y) &=&  \frac{1}{2}\left(I(x,y+1)-I(x,y-1)\right).  \label{eq.02} %\\
\end{eqnarray}
In order to decrease the noise at CFD after JPEG/JPEG2000 compression it is
proposed a convolution of the image luminance
$\left(I(x,y)|\ (x,y)\in\{0,1,\ldots,n_x\}\times\{0,1,\ldots,n_y\}\right)$
with a two-dimensional Gaussian-wise filter having pulse response:
$$h(i,j) = \left\{\begin{array}{ll}
 \frac{1}{2\pi\sigma^2}\,\mbox{\rm exp}\left(\frac{\left(i-\frac{n}{2}\right)^2 + \left(j-\frac{n}{2}\right)^2}{2\sigma^2}\right) & \mbox{ if }1\leq i,j\leq n \\
 0 & \mbox{ otherwise }
\end{array}\right.$$
where $\sigma^2$ characterizes the filtering parameter and $n$ is the size of the Gaussian-wise window. After the two-dimensional convolution $h ** I$ we get
\begin{equation}
\tilde{I}(x,y) = \sum_{i=0}^{n-1} \sum_{j=0}^{n-1}h(i,j)\,I(x-i,y-j). \label{eq.03}
\end{equation}
The corresponding CFD's $\left(\tilde{\delta_x}(x,y)\right)_{x,y}$, $\left(\tilde{\delta_y}(x,y)\right)_{x,y}$ are obtained as in~(\ref{eq.01})-(\ref{eq.02}) with replacing $I(x,y)$ by $\tilde{I}(x,y)$.
Let us define
$$%\begin{equation}
\tilde{\delta}(x,y) = \sqrt{\tilde{\delta_x}(x,y)^2+\tilde{\delta_y}(x,y)^2} %\label{eq.04}
$$%\end{equation}
and let us consider the matrix ${\bf G} = \left[\tilde{\delta}(x,y)\right]_{x,y}$ (It is worth to note that the computation of $\tilde{I}(x,y)$ by~(\ref{eq.03}) can be simplified when the CFD's are presented as a trivial convolution). The elements of the matrix ${\bf G}$ are changing slightly after JPEG compression but unfortunately the size of this matrix is still very large to be embedded into DI. In order to decrease the size of this matrix it is proposed to perform the so called \emph{average downsampling} technique~\cite{xv} with integer parameters $s$ and $t$, corresponding divisors of $n_x$, $n_y$, for horizontal and vertical directions: $\forall (k,m)\in\left\{1,\ldots,\frac{n_x}{s}\right\}\times\left\{1,\ldots,\frac{n_y}{t}\right\}$ as follows
\begin{equation}
d(k,m) =  \frac{1}{st}\sum\{\tilde{\delta}(i,j)|\ s (m -1) < i \leq s m\ \& \ t (k -1) < j \leq t k\ \} \label{eq.05}
\end{equation}
We call matrix ${\bf D} = \left[d(k,m)\right]_{k,m}$ the \emph{image feature matrix}. Then, let us quantize the values $d(k,m)$ with some step $\Delta\in\R$ as
\begin{equation}
d_{\Delta}(k,m) = \left\lfloor\frac{d(k,m)}{\Delta}\right \rfloor  + 1 \label{eq.06}
\end{equation}
where $\lfloor\cdot\rfloor$ is the floor map. For simplicity, we will present these values as a linear array,
$$d_{\Delta}=\left(d_{\Delta}(i)\right)_{i=1}^{\frac{n_xn_y}{st}}.$$
Now, it would be possible to authenticate the tested image $\left(\tilde{I}(x,y)\right)_{x,y}$, given the embedded vector $d_{\Delta}$ and the corresponding vector $\tilde{d_{\Delta}}$ calculated for the image $\left(\tilde{I}(x,y)\right)_{x,y}$. Then, the following condition has been taken for the authentication rule
\begin{equation}
\left(\tilde{I}(x,y)\right)_{x,y} \mbox{is authentic}\ \Longleftrightarrow\ \max_i|\tilde{d_{\Delta}}(i) - d_{\Delta}(i)| \leq 1 .  \label{eq.07}
\end{equation}
But the use of the authentication rule~(\ref{eq.07}) is inconvenient for two reasons. Firstly, the size of the authenticator $d_{\Delta}$ is rather big to be embedded into the image without significant corruption. Secondly, any adversary might be able to forge the authentication process because no cryptographic technique was used here. In order to avoid these two defects, it is necessary to hash the feature vector $d_{\Delta}$ and to obtain its digital signature. On the other hand, hashing the vector $d_{\Delta}$ after its corruption by JPEG compression results in error expansion. In order to recover $d_{\Delta}$, after jumps of their coordinates in at most one quantization level, it is possible to use the so called \emph{3-bit quantization} technique~\cite{xvi} briefly considered below.

Let us introduce an \emph{auxiliary perturbation vector} $p$ of dimension $N = \frac{n_xn_y}{st}$ where the \emph{i}-th coordinate contains three bits $p_{1i},p_{2i},p_{3i}$ computed as follows:
\begin{eqnarray}
(p_{1i},p_{2i}) &=& [d_{\Delta}(i)\bmod 4]_2  \label{eq.08} \\
p_{3i} &=& \left\{\begin{array}{cl}
 1 & \mbox{ if } d(i) \in\left[a_i, b_i\right) \\
 0 & \mbox{ if } d(i) \in\left[b_i, a_{i+1}\right) %\\
\end{array}\right.  \label{eq.09} %\\
\end{eqnarray}
with $a_i= \Delta\,d_{\Delta}(i)$, $b_i= \Delta\,\left(d_{\Delta}(i)+\frac{1}{2}\right)$, and $[\cdot]_2$ the binary representation of the argument integer.

Then, the vector $d_{\Delta}$ can be hashed. The obtained hash is signed with the use of cryptographic DS and then this DS is embedded jointly with the auxiliary perturbation vector $p$ into the image. Verification of DS is performed by conventional cryptographic methods, where it is necessary only to recover the feature vector $\tilde{d_{\Delta}}$, corrupted possibly by JPEG/JPEG2000 compression of the original feature vector $d_{\Delta}'$. This can be performed as follows~\cite{xvi}
\begin{equation}
d_{\Delta}'(i) = \left\lfloor\frac{d'(i)}{\Delta}\right\rfloor \label{eq.10}
\end{equation}
where
$$%\begin{equation}
d'(i) = \left\{\begin{array}{ll}
 \tilde{d}(i) + \Delta & \mbox{ if }\alpha_i =0\ \&\ \tilde{p}_{3i} = 0 \\
 \tilde{d}(i) + \Delta & \mbox{ if }\alpha_i =0\ \&\ \tilde{p}_{3i} = 1\ \&\ p'_{3i} = 1 \\
 \tilde{d}(i) - \Delta & \mbox{ if }\alpha_i =1\ \&\ \tilde{p}_{3i} = 1 \\
 \tilde{d}(i) - \Delta & \mbox{ if }\alpha_i =1\ \&\ \tilde{p}_{3i} = 0\ \&\ p'_{3i} = 0 \\
 \tilde{d}(i)  & \mbox{ otherwise }
\end{array}\right. %\label{eq.11}
$$%\end{equation}
and
\begin{equation}
\alpha_i = \left\{\begin{array}{ll}
 0 & \mbox{ if }[p'_{1i}p'_{2i}]_{10} = ([\tilde{p}_{1i}\tilde{p}_{2i}]_{10}-1)\bmod 4 \\
 1 & \mbox{ if }[p'_{1i}p'_{2i}]_{10} = ([\tilde{p}_{1i}\tilde{p}_{2i}]_{10}+1)\bmod 4 \\
 2  & \mbox{ otherwise }
\end{array}\right. \label{eq.12}
\end{equation}
where $[\cdot]_{10}$ is the decimal representation of the binary integer;  $(\tilde{p}_{1i},\tilde{p}_{2i},\tilde{p}_{3i})$ are the three bits of each entry $\tilde{p}_i$ of the perturbation vector $\tilde{p}$ extracted as a WM and $(p'_{1i},p'_{2i},p'_{3i})$ are obtained from the perturbation vector $p'$ calculated by~(\ref{eq.08}),~(\ref{eq.09}) given by the corrupted image  $\left(\tilde{I}(x,y)\right)_{x,y}$; $\tilde{d}(i)$ is the \emph{i}-th element of the feature vector given by~(\ref{eq.06}) and the image $\left(I(x,y)\right)_{x,y}$ is the original one before recovering.

It has been proven in~\cite{xvi} that the feature vector $\tilde{d}_{\Delta}$ can be recovered exactly by~(\ref{eq.10})--(\ref{eq.12}) if the extracted auxiliary perturbation vector $\tilde{p}$ is correct and the rule~(\ref{eq.07}) is achieved. This rule will be valid if a corruption of the quantized feature vector entries $\tilde{d}_{\Delta}(i)$ have transitions to at the most one neighbor quantization level. Hence, the proposed authentication method will be tolerant to JPEG compression of DI if the quantization step was chosen in such a way that the last requirement holds with the high probability. In the next section the methods of embedding and extraction providing acceptable error probability of both feature vector signature and auxiliary perturbation vector are considered.

\section{Embedding and extraction algorithms for the proposed authentication method} \label{sc.III}

In order to embed both the feature vector $d$ and the auxiliary perturbation vector $p$, the following properties are necessary for the embedding algorithm:
\begin{itemize}
\item  robustness to JPEG/JPEG2000 compression;
\item  sufficient capacity to embed both $d$ and $p$;
\item  lower computational complexity; and
\item  keeping a good DI quality just after embedding.
\end{itemize}

\begin{figure}
\centering
\includegraphics[width=2in]{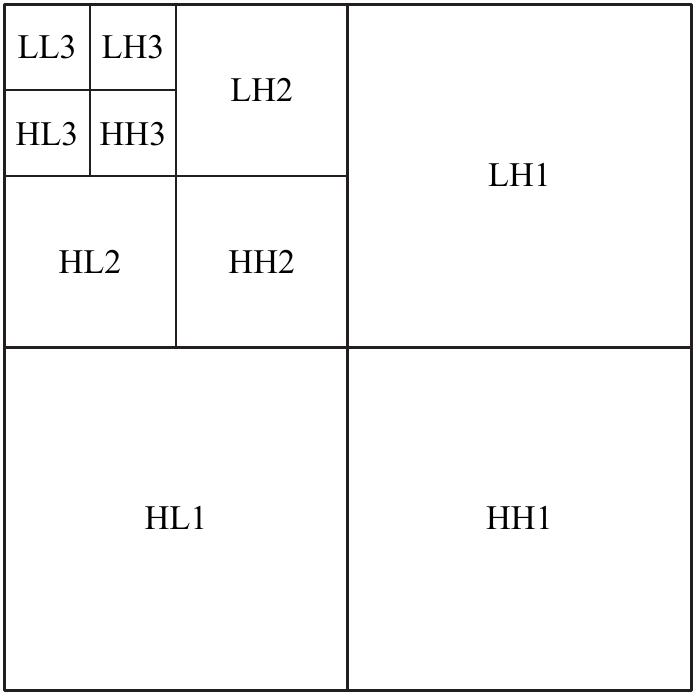}
\caption{Areas of the 3-level two-dimensional HWT with conventional
notations.}
\label{fg.01}
\end{figure}
\begin{figure*}[h!]
\centering
\includegraphics[width=0.85\textwidth]{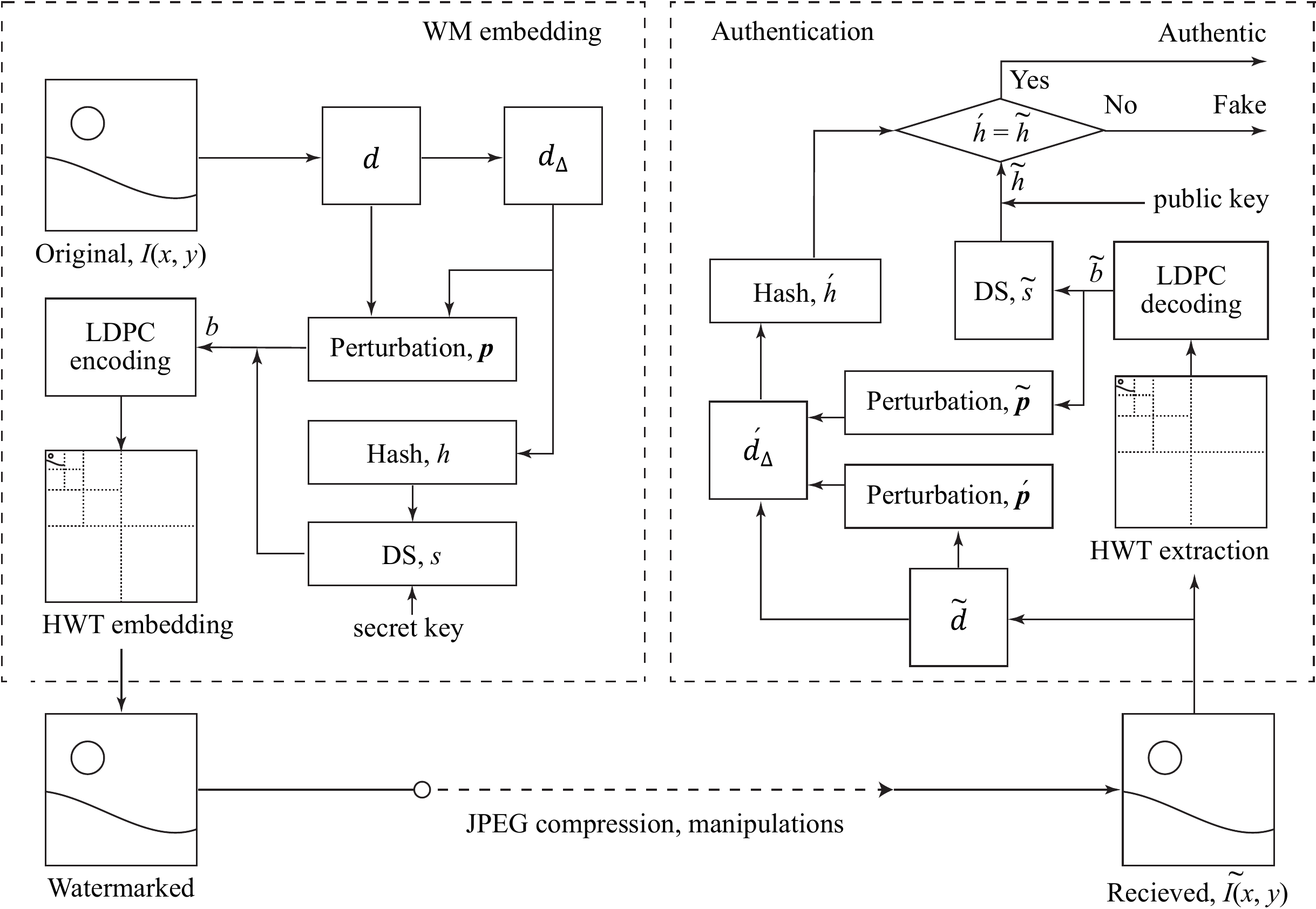}
\caption{The general scheme of embedding and extraction algorithms for the proposed semi-fragile authentication method tolerant to JPEG compression.}
\label{fg.02}
\end{figure*}

Taking into account the requirements presented above, the embedding algorithm based on coefficients quantization of 3-level discrete {\em Haar Wavelet Transform} (HWT)~\cite{xvii} was selected. It was chosen only LH3 and HL3 submatrices for WM embedding because of their robustness to small noises that can be introduced by JPEG compression. However, as our experiments show, coefficients of LL3 having more evident influence on visual image quality after embedding should be omitted. Let us assume for simplicity that the DI is square, of order $2^m\times 2^m$. (We note that if the image I is not represented by a square matrix then it can be padded with zero elements). According to~\cite{xvii}, two-dimensional forward and inverse HWT of the square image luminance values $(2^m\times 2^m)$-matrix I of the can be found as:
$$S = H_mIH_m^T\ \ ,\ \ I=H_m^TSH_m,$$
where $S$ is the matrix of the HWT coefficients, and the super index $T$ denotes matrix
transposition. The recurrent relations~\cite{xvii}
$$H_0 = [1]   \ \ , \ \
H_m = \frac{1}{\sqrt{2}}\left[\begin{array}{rr}
H_{m-1} &  H_{m-1} \\  H_{m-1} & -H_{m-1}
\end{array}\right]$$
determine the $(2^m\times 2^m)$-Haar single level matrices $H_m$. The next level of HWT can be obtained if, instead of the original image $I$, the $(2^{m-1}\times 2^{m-1})$-submatrix of HWT approximation coefficients is used. Figure~\ref{fg.01} shows the 3-level HWT coefficients submatrices with conventional notations.

As mentioned above, we chose only LH3 and HL3 coefficients as they represent low-frequency components of the image and have explicit robustness to the distortions introduced by JPEG/JPEG2000 compression, see Figure~\ref{fg.01}. Such approach allows to minimize DI corruption after embedding. The general schemes of embedding and extraction algorithms are presented in Figure~\ref{fg.02} with correspondent notations.
The quantized feature vector $d_{\Delta}$ is hashed and signed by some standard cryptographic algorithms~\cite{i} giving strong digital signatures $s$. Next, this DS and perturbation vector $p$ is concatenated into one binary string $b$. In order to increase efficiency of authentication data transferring in the presence of corrupting noise {\em Low-Density Parity-Check} (LDPC) code~\cite{Gallager} was applied. Encoded block $b_e$ represented by digits $b_{e_k}$ is embedded into the coefficients $S_k$ belonging to HWT areas HL3 and LH3 (see Figure~\ref{fg.01}) by the following rule:
\begin{equation}
\tilde{S}_k = \left\{\begin{array}{ll}
 \gamma\left(\left[\frac{S_k}{\gamma}\right] + \frac{1}{4}\right) & \mbox{ if }b_{e_k}=1 \vspace{2ex}\\
 \gamma\left(\left[\frac{S_k}{\gamma}\right] - \frac{1}{4}\right) & \mbox{ if }b_{e_k}=0
\end{array}\right. \label{eq.13}
\end{equation}
where $\gamma$ is a quantization interval of HWT coefficients, $[\cdot]$ is the nearest integer of a real number, and $\tilde{S}_k$ is the coefficient after embedding the bit $b_{e_k}$.

In order to verify that DS is authentic, see Figure~\ref{fg.02}, it is necessary firstly to take a decision $\tilde{b}_{e_k}$ regarding the digits of the binary string $b_e$ using the decision rule
\begin{equation}
\tilde{b}_{e_k} = \left\{\begin{array}{ll}
 1 & \mbox{ if }\tilde{S}_k -\gamma\left[\frac{\tilde{s}_k}{\gamma}\right] \geq 0 \vspace{2ex}\\
 0 & \mbox{ if }\tilde{S}_k -\gamma\left[\frac{\tilde{s}_k}{\gamma}\right] < 0
\end{array}\right. \label{eq.14}
\end{equation}
where $\tilde{S}_k$ are the coefficients $S_k$ into HWT areas HL3, LH3 that might be corrupted by some DI transforms. Decoding of received code word $\tilde{b}_e$ was implemented with technique based on {\em iterative belief propagation} technique~\cite{Guruswami}.

Next, it is necessary to extract the elements of the perturbation vector $\tilde{p}$ from decoded data using~(\ref{eq.14}) and the vector $p'$ calculated directly from the image by~(\ref{eq.08}),~(\ref{eq.09}), and then recover $d'_{\Delta}$, given the vectors $\tilde{d}_{\Delta}$, $\tilde{p}$ and $p'$. Then the recovered vector $d'_{\Delta}$ is hashed to $h'$ and compared with the hash $\tilde{h}$ obtained from the DS $\tilde{s}$ with the use of the corresponding public key. If $h'=\tilde{h}$ then DI is recognized as authentic, otherwise it is assumed as fake one.

\section{Simulation results and optimization of the parameters for the proposed authentication system} \label{sc.IV}

First of all, it is necessary to investigate the sensitivity of the authentication system to JPEG/JPEG2000 compression. It will be tolerant to such compression if the rule~(\ref{eq.07}) holds. We have selected 50 different $512\times 512$ DI having varied content, textures and so forth. Then the HL3 and LH3 areas of HWT contains $2\times (2^6)^2 = 2^{13}= 8192$ coefficients. Every of them allows to embed one bit by the rule~(\ref{eq.13}).

In order to provide some redundancy, the length of the feature vector $d$ was taken as $2^{10}= 1024$ corresponding to the sizes $2^5\times 2^5 = 32\times 32$ of the matrix ${\bf D}$. This requires to take parameters $s=t= 2^4=16$ in~(\ref{eq.05}). In such a way, the auxiliary perturbation vector $p$ has $3\times 2^{10} = 3072$ bits length. As hash function, it was used the standard SHA-2~\cite{i} and the DS algorithm based on RSA cryptosystem with length of modulo 1024 bits. Then, the total size of the embedded bits is  $3072 + 1024 = 4096$ bits. Given the total number of HWT coefficients in embedding domains it was chosen the $(8192, 4096)$-LDPC code to achieve appropriate error correction.

\begin{figure}
\centering
\includegraphics[width=6in]{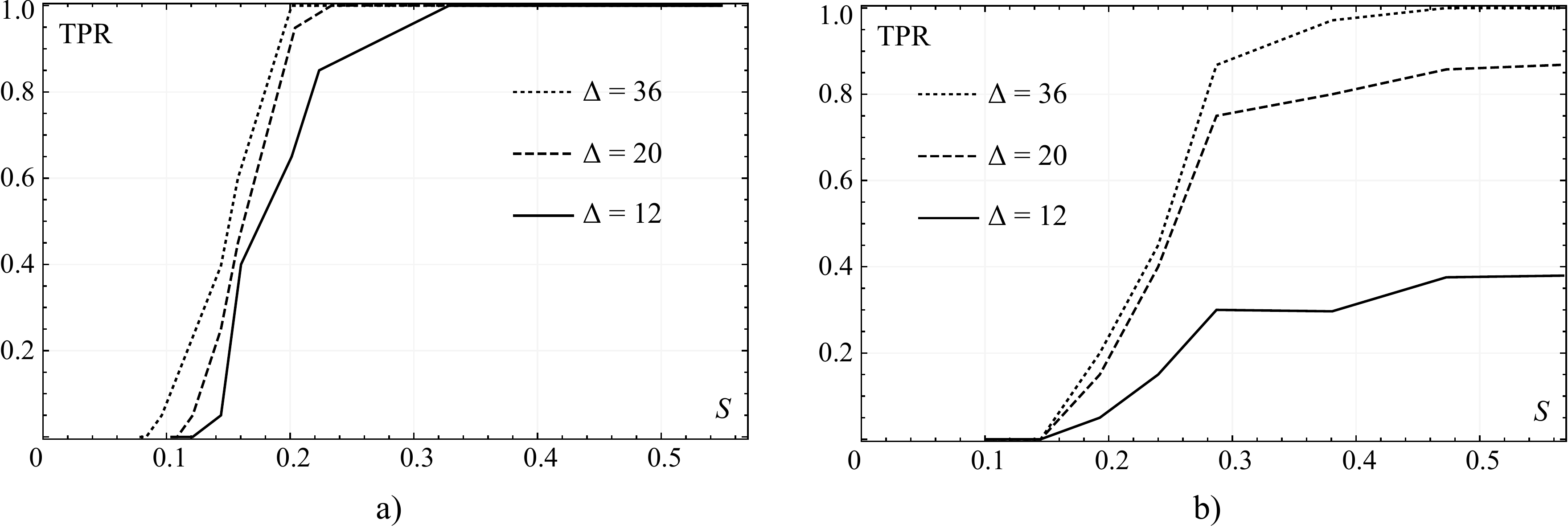}
\caption{Dependences of the {\em True Positive Rate} (TPR) against {\em space savings} $S$ given by: a) - JPEG, b) - JPEG2000, with different quantization steps $\Delta$.}
\label{fg.03}
\end{figure}

After the selection of the main parameters, it was performed an investigation of the authentication system efficiency.  Figure~\ref{fg.03} shows the dependences of the {\em True Positive Rate} (TPR) against {\em space savings} $S = M_c / M_0$, where $M_0$, $M_c$ are sizes of the image before and after compression by means of: a) - JPEG, b) - JPEG2000, with different quantization steps $\Delta$ used in the formation of $d_{\Delta}(k,m)$ by~(\ref{eq.06}).
We can see from Figure~\ref{fg.03} that the greater is the {\em Compression Rate}, the better is image authentication method resistant to JPEG compression. However, due to the features of JPEG2000 that makes an image more blurred, proposed method has greater robustness to conventional JPEG compression.

\begin{figure}
\centering
\includegraphics[width=3.2in]{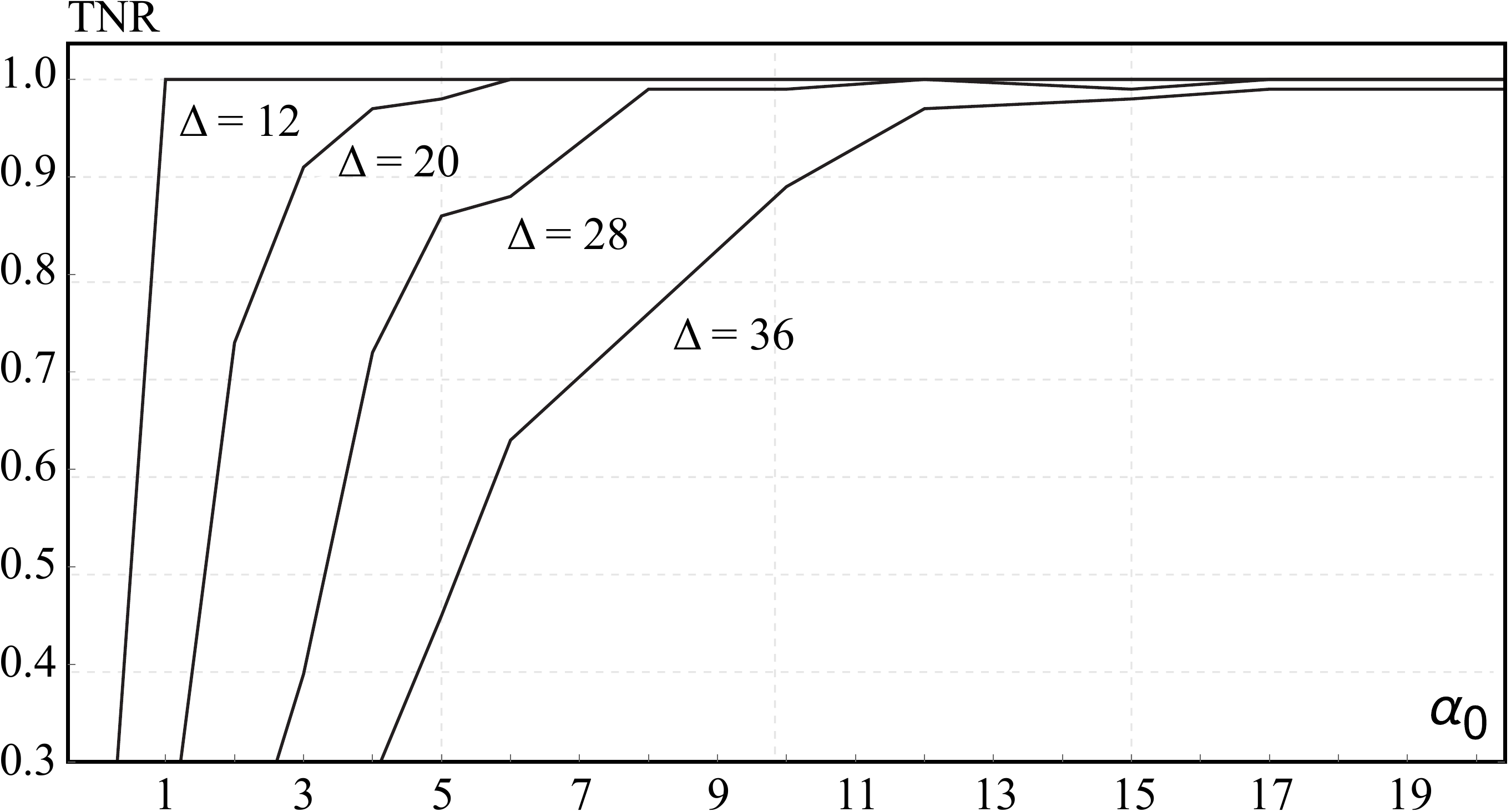}
\caption{Dependence of authentication TNR against the area size $a_0$ for different quantization steps $\Delta$ for feature vector coordinates.}
\label{fg.04}
\end{figure}

The strongest requirements should be formulated for the opportunity to detect all image pixel modifications except for JPEG/JPEG2000 compression, for instance, some random modifications or malicious attacks intended to compromise the original image, for the thing, changing of car plate numbers for DVR systems, or fingerprints and photos of criminals in police offices. It is a trivial problem for exact authentication, provided that the cryptographic components, namely hash function and DS were selected in an appropriate manner. But it is a relevant problem for semi-fragile authentication because in this case some modifications may not be detected. In order to verify such opportunity for the proposed system we arranged the following experiment. It was selected a truly randomly square $(a_0\times a_0)$-pixel areas and inside of these areas it was chosen truly random luminance of pixels. The number of these areas was taken at least 50 for each of the images and the number of different typical images as 100. The results of testing are presented at Figure~\ref{fg.04}, where a dependency of \emph{True Negative Rate} (TNR) is showed as a function of areas size $a_0$ depending on quantization step $\Delta$ of the feature vector coordinates.

From Figure~\ref{fg.04}, it can be seen that, in line with intuition, the less is a quantization step $\Delta$, the more probably to detect small image modifications.

\begin{figure}
\centering
\includegraphics[width=3.2in]{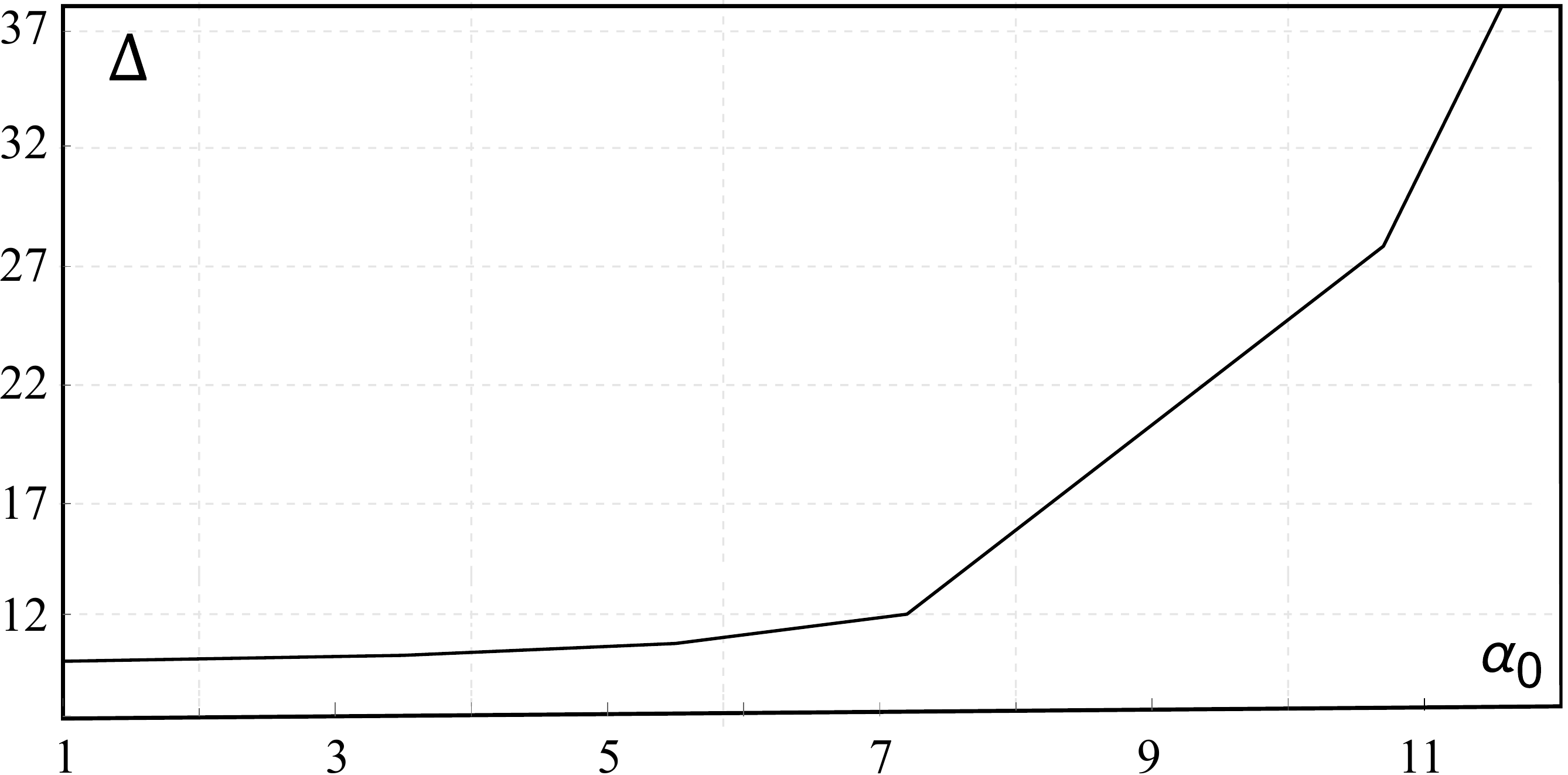}
\caption{Dependence of the requested quantization step values $\Delta$ against the modification area size $a_0$ given TNR$\geq 0.95$.}
\label{fg.05}
\end{figure}

In Figure~\ref{fg.05} there is presented a curve showing a dependence of the requested values of quantization steps $\Delta$ against the size of modification area $a_0$ given a TNR of value at least 0.95.

By summarizing the experimental results, we can conclude that the proposed authentication method is tolerant to JPEG/JPEG2000 compression with parameter $CR\leq 0.3$ providing simultaneously TNR$\geq 0.95$ for modification area size $a_0\geq 8$.

\begin{figure}
\centering
\includegraphics[width=6.8in]{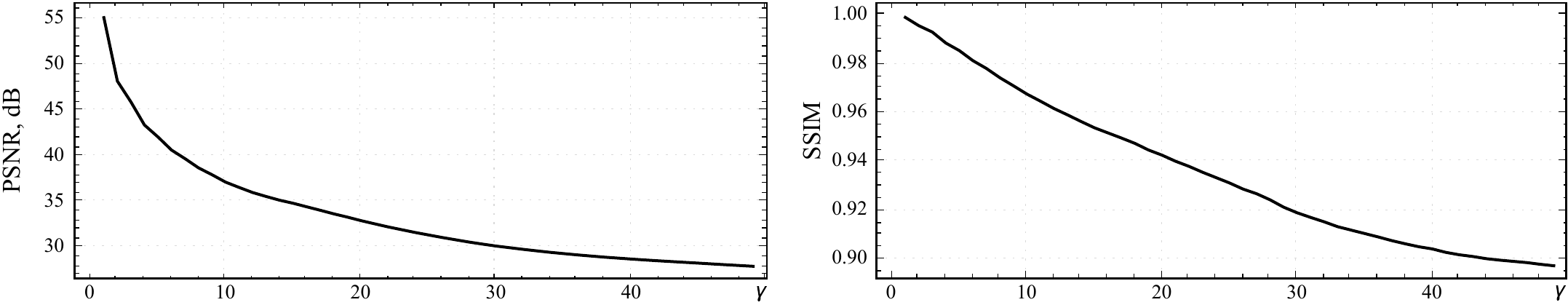}
\caption{The dependences of image quality assessments PSNR and SSIM just after WM embedding against HWT coefficients quantization step $\gamma$.}
\label{fg.06}
\end{figure}

Image quality of DI just after WM embedding is also very important criterion of authentication system efficiency. We evaluate both {\em Peak Signal-to-Noise Ratio} (PSNR) and {\em Structural Similarity Index Measure} (SSIM)\cite{metrics} as they very common in the literature and image quality assessments can be compared with other methods. In Figure~\ref{fg.06} it is presented the curves of image quality assessments PSNR and SSIM (in line with proposals~\cite{xix}) depending on the quantization step $\Delta$. We can see from this figure that the greater is $\Delta$, the worse is the visual comprehension of the images. On the other hand the proposed system requires to keep $\Delta$ to be not very small in order to WM be tolerant to JPEG compression.

\begin{figure}
\centering
\includegraphics[width=5in]{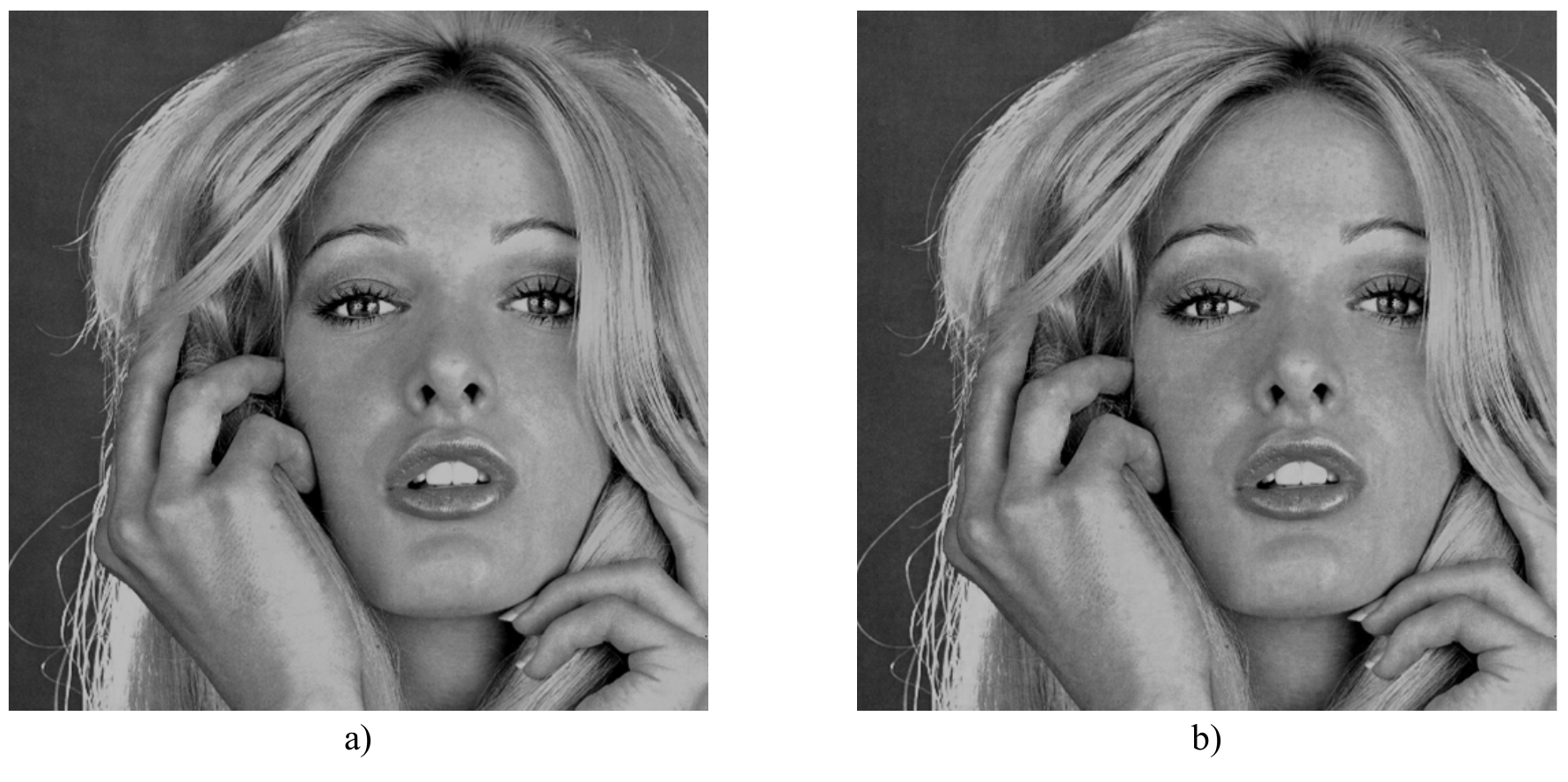}
\caption{Original (a), and its watermarked version (b) of the image given by quantization interval of HWT coefficients  $\gamma = 40$.}
\label{fg.07}
\end{figure}

In Figure~\ref{fg.07} it is displayed the visual effect of WM embedding for some chosen WM system parameters. There is no opportunity to find some differences between images (a) and (b). However, it is worth to note that reliable detection of the image modification has a greater importance than false detection after JPEG/JPEG2000 compression, because in the last case an error can easily be recognized, whereas the authenticated image content corruption may lead to fatal consequences.

\begin{figure}
\centering
\includegraphics[width=6in]{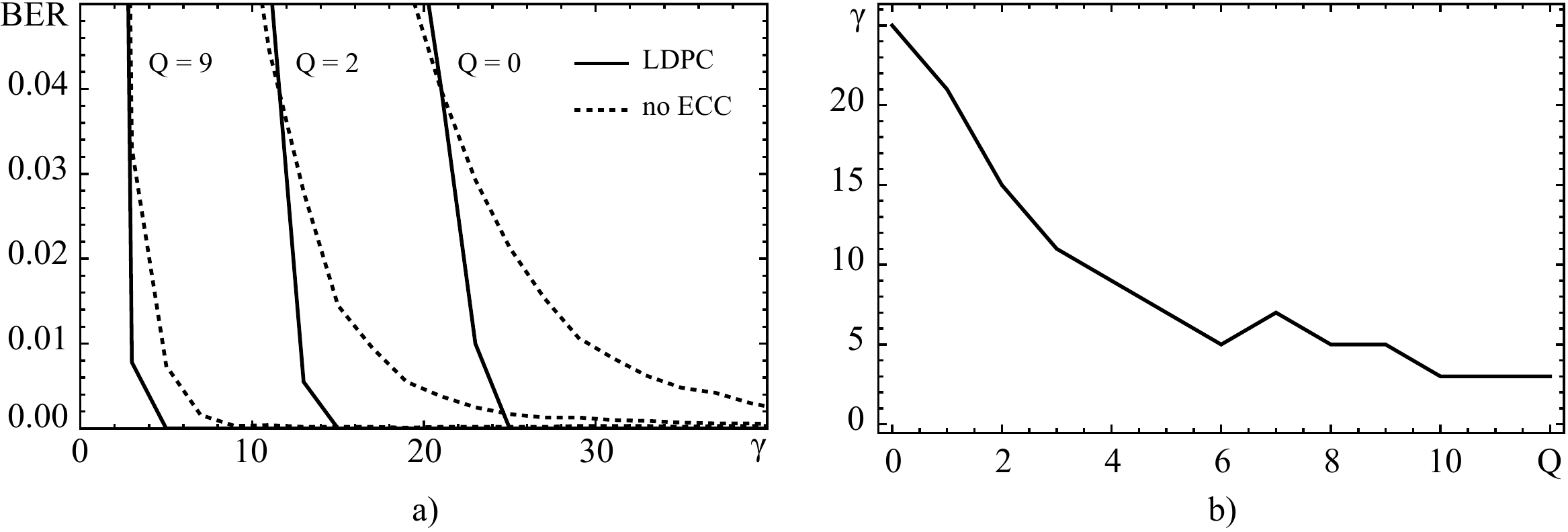}
\caption{(a) Dependencies of bit error rates (BER) against quantization step $\gamma$ for HWT coefficient depending on different JPEG compression factors $Q$. (b) Dependence of the quantization intervals $\gamma$ on JPEG compression quality factor $Q$ given the condition BER$=0$.}
\label{fg.08}
\end{figure}

It is obvious that for valid authentication system operation, the error probability after decoding of WM should be equal to zero even after image compression. Figure~\ref{fg.08} serves in order to specify this problem. Figure~\ref{fg.08} (a) shows a dependencies of {\em bit error rates} (BER) against quantization step $\gamma$ for HWT coefficients depending on different values of JPEG compression factor $Q$. It can be seen that there are values of $\gamma$ leading to BER$=0$ in case when LDPC coding is applied, whereas on the contrary in case of watermarking without error correction code (ECC) BER is mostly non-zero. Figure~\ref{fg.08} (b) presents the dependence of the quantization intervals $\gamma$ on JPEG compression quality factor $Q$ given the condition BER$=0$.

From Figure~\ref{fg.08} it can be seen that the selection of the quantization interval $\gamma$ equal to 10 provides a resistant authentication method to JPEG compression with quality factor $Q\geq 4$ and quality assessments $\mbox{\rm PSNR}\geq 40$, $SSIM > 0.98$ that can be assumed as acceptable values.

\section{Conclusion} \label{sc.V}

A semi-fragile authentication system has been proposed in this paper that can be considered as an improved selective authentication presented previously~\cite{xiii}. The main idea to solve this problem was to use image central finite differences as the property functionals of the image and next to apply 3-bit hash quantization providing a recovering of hash function even after jumps it arguments due to JPEG/JPEG2000 compression.

The 3-level discrete HWT was proposed in order to embed authentication data into DI, namely into massive of HWT coefficients belonging to HL3 and LH3 areas. Experimental investigation showed that proposed authentication method provides a good reliability to verify image quality authenticity even after JPEG/JPEG2000 compression with CR $\leq 0.3$ and simultaneously an opportunity to recognize even small content image modifications and image quality assessments $\mbox{\rm PSNR}\geq 40$ and $SSIM > 0.98$ just after WM embedding.

The proposed authentication system has an opportunity to trade-off the main criteria, such as BER, \emph{Q}, PSNR, SSIM, TNR, and TPR, one to another and hence to choose the most important characteristics that are better than other with designer's point of view. With computational point of view the proposed method is simple enough. It requires less than about 0.1 sec for WM embedding and 0.2 sec for WM verification on computer with Intel Quad Core i5, 2.5 GHz processor.

It is worth to note that our method to form the feature vector after small modification is also tolerant to such image transforms as changing of size, luminance and contrast, that can be taken as practically full set of acceptable manipulations for content authentication.

However embedding and extraction algorithms after such transform belong still to open problem. Authors hope to solve them in the nearest future.

\bibliographystyle{IEEEtran}
\bibliography{IEEEabrv,references}

% That's all folks!
\end{document}